\documentclass[10pt, sigconf]{acmart}
\usepackage{booktabs} 
\usepackage{mathtools}
\usepackage{bm}
\usepackage{algorithmic}
\usepackage{graphicx}
\usepackage{textcomp}
\usepackage{xcolor}
\usepackage{subcaption}
\usepackage{diagbox}
\usepackage{tabulary}
\usepackage{mathtools}

\setcopyright{rightsretained}

\acmDOI{}

\acmISBN{}

\acmConference[]{}{}{}
\copyrightyear{2021}

\begin{document}

\date{}

\title{\Large \bf Bandcoin: Using Smart Contracts to Automate\\
Mobile Network Bandwidth Roaming Agreements}


\author{ \small Thomas Sandholm and Sayandev Mukherjee} 
\affiliation{
  \institution{Next-Gen Systems, CableLabs}
}

\begin{abstract}
We propose a new way to share licensed spectrum 
bandwidth capacity in mobile networks between 
operators, service providers and consumers using 
blockchain-based smart contracts.
We discuss the foundational building blocks in
the contract as well as various extensions to
support more advanced features such as bulk
purchases, future reservations, and various
auction mechanisms. Furthermore, we demonstrate 
how the system can be implemented with an 
open-source, permissioned Enterprise
blockchain, Hyperledger Sawtooth. We
show that our smart contract
implementation can improve
blockchain transaction performance,
by approximately four orders of magnitude compared to serial transactions
and one order of magnitude compared to parallell transactions, using PKI-driven
bulk purchases of mobile access grants, paving the way
for fully automated, efficient, and fine-grained roaming agreements.
\end{abstract}

\maketitle

\section{Introduction}\label{sec:introduction}
\subsection{The challenge of efficient spectrum allocation}
Mobile cellular network providers today face increasing
challenges when it comes to acquiring licenses to operate on
dedicated frequency bands in large coverage areas. Winning
a license in national auctions typically also comes with
strings attached to build out network infrastructure and show
a high level of utilization to serve the general public.
The high cost of provisioning the network is translated
into more expensive and rigid contracts to lock in consumers
or long-term partnerships with service providers, as well as limited
competition and innovation.

The end-result is poor quality of experience for consumers, in particular
highly mobile ones, and inefficient resource utilization.
The additional spectrum bandwidth introduced in newer
specifications such as 5G, exacerbate rather than mitigate this problem
as the existing roaming and bandwidth sharing mechanisms are not up to the
task of sharing the new opportunities efficiently, mostly due to the rigid
contractual setup of roaming agreements.

There is a trend to design more dynamic database mediated allocations
of spectrum as evident in efforts like CBRS SAS~\cite{sohul2015}. However, even in that work
there is a centralized mediator which allows long term spectrum
licenses to be acquired in smaller scale spectrum auctions (e.g. county level), 
and most of the
dynamic coordination is focused on letting unlicensed traffic use up
incumbent and licensed bands if not in use, as opposed to redesigning
the actual license acquisition process. Thus it inherits many of
the inefficiencies of the national spectrum auctions, just on a smaller
scale.

\subsection{Markets for efficient allocation of scarce spectrum}
We know from economics that markets are an efficient way to allocate resources in such a way that both buyers and sellers are satisfied with the value they receive from transactions in such a market.  Wireless spectrum is presently mostly spoken for due to long-term licenses that exemplified the market for a scarce resource (spectrum) with the government as the seller and mobile network operators (MNOs) as the buyers.  However, this is not a dynamic market because transactions (i.e., license auctions) occur infrequently and licenses are allocated for fixed blocks of spectrum for durations of several years.  Instead, one can conceive of a different, much more dynamic market, namely one where the end-user is the buyer, the seller is any entity that can allocate a fine-grained amount of bandwidth to the buyer for a short interval that may be just a few seconds or minutes long, and transactions between sellers and buyers are fast and frequent. In this market, the rapid turnover in use of any given band of spectrum ensures that spectrum resources are fully utilized.  Although the sellers of spectrum are initially going to be the incumbent holders of spectrum licenses (i.e., the MNOs who acquired that license in a spectrum auction), it is conceivable that in the long run the method of spectrum allocation via long-term licenses goes away and the seller of spectrum is the government, which will sell spectrum for short durations (ranging from seconds to minutes) directly to other entities that may not necessarily be service providers themselves (see below).

To maximize the efficiency of spectrum allocation and usage we believe there
are three key dimensions where granularity needs to be reduced:
\begin{itemize}
\item{{\bf Time.} Provide shorter-term contracts. Not years, weeks, or even days, but hours, minutes or seconds.}
\item{{\bf Location.} Provide contracts for smaller geographic regions. Not national, state, or county, but radio range, or street block.}
\item{{\bf Spectrum chunks.} Provide more flexibility in the bandwidth allocated in each contract. An operator may have a 
license for 100\,MHz but could then dole that out to consumers in any amount from 1\,MHz to 100\,MHz based on demand.}
\end{itemize}

It is a consequence of a dynamic market that there is a role to play for an intermediary between the ``owners'' of spectrum (the MNOs today) and the end-users of that spectrum.  This intermediary plays the role of the service provider to the end-users, but is a bulk customer of the spectrum owners (the MNOs).  There is a role for this intermediary because MNOs would prefer not to have to deal with millions of transactions representing sales to individual end-users every few seconds or minutes, and nor would the end-users necessarily prefer dealing with multiple MNOs in order to procure bandwidth to use for a short duration.

Note that each use by an end-user of a block of licensed spectrum with the permission of the licensee corresponds to a ``contract'' wherein the licensee gives that end-user access to the bandwidth corresponding to that block of spectrum in return for a payment from that end-user either directly or indirectly via a service provider which collects payment from a number of end-users.  Addressing the additional flexibility of a market supporting all the above makes it apparent that contract negotiation, establishment, accounting, and verification all need to be automated in order to handle the high volume of transactions that will be generated in such a dynamic market. Such contracts, validation, billing, accounts payable/receivable etc. can all be automated using the technology of ``smart contracts,'' which can therefore be used to handle all transactions in such a market.  Today, such smart contracts are usually implemented on a blockchain platform and are used in everything from
online shopping to house mortgage lending.  

\subsection{Outline of this paper}
Our goal is not simply to design an open market of spectrum bandwidth exchange based on smart contracts, 
but to design it in such a way that operators, service providers as well as users have an incentive
to participate, at the same time making it as easy as possible for them to do so, all without requiring massive investment in infrastructure
beyond what is already deployed in the field, and without requiring protocol changes that would deprecate all current devices.
Furthermore, we would like to facilitate operator-to-operator, operator-to-service provider, operator-to-consumer, as well
as service provider-to-consumer exchanges of spectrum bandwidth, backed by a readily usable mobile network.

The rest of the paper is organized as follows: we discuss related work, the current roaming mechanism, and blockchain
fundamentals. We then describe the basic building blocks of our implementation as well as extensions for more sophisticated
auction mechanisms. Finally, we present details about our implementation and conclude with lessons learned.

\section{Related Work}\label{sec:relatedwork}

\subsection{Cognitive Radio}
In \cite{chapin2007}, in the context of cognitive radios
and Dynamic Spectrum Access (DSA), the authors argue
that although technical advances have been achieved
to enable DSA, the spectrum market also needs to
evolve to support commercial use. Their problem formulation 
is different from the present work in that in~\cite{chapin2007}, the market
governing spectrum licenses is intended for multiple
spectrum operators to coordinate usage dynamically,
i.e. base stations deciding on which band is best
to operate on considering other interfering base stations.
Our approach is more end-user focused in that we allow
a spectrum operator who already as obtained the
license to operate and provide cellular licensed band
service to admit non-subscribers to their network, similar
to how roaming is done, governed by an open bandwidth
market. Nevertheless, their analysis on why DSA and
wireless spectrum markets are beneficial applies to our
approach as well. The authors argue that
more spectrum sharing leads to cheaper spectrum
access that in turn enables increased competition
among mobile communication services, which necessitates 
more innovation and allows smaller operators to thrive, 
and ultimately leads to overall more affordable 
mobile services. Driving down transaction cost is stated
as a primary challenge, and it is also the focus of
our work. The analysis by \cite{chapin2007} pre-dated
but predicted, and recommended the regulatory frameworks that later appeared in the FCC CBRS (Citizen Broadband Radio Service) regulations.

In \cite{xu2010}, the authors propose a secondary market auction design
for base stations to share spectrum (as channel holdings) in order to leverage temporal, spatial and channel variation in user demand. A dynamic double auction, akin to the stock market and similar to our proposed Vickrey auction, is applied.
The authors furthermore analytically prove the truthfulness and 
asymptotic efficiency in maximizing spectrum utilization.
Our market is different due to end-users, not base stations, 
participating directly in the auctions. Our market is also more of
a bandwidth market as we sell not only a lease to be licensed to
operate on a spectrum but a full grant to access a mobile network
as your home-PLMN over a limited time.

\subsection{CBRS}
The regulatory framework of Citizen Broadband Radio Service (CBRS)~\cite{CBRS} allows primary, secondary and incumbent spectrum
owners to share bandwidth dynamically through
a Spectrum Access System (SAS)~\cite{sohul2015} that interacts with environment sensing services as well as base stations requesting
spectrum to mitigate interference on both licensed and unlicensed bands. Incumbents are typically federal radar
applications along the US coastlines. 
In \cite{yrjola2017}, the author argues that blockchains
could help with the CBRS use case in terms of 
building trust, consensus and reducing transaction cost.
One interesting aspect of CBRS is that the FCC allows
and encourages Priority Access
License (PAL) holders to resell and lease out their
spectrum rights on secondary markets. Key benefits
offered by blockchains include reduced transaction cost,
which we recall was one of the promises of DSA.
A large variety of CBRS use cases of blockchains are
proposed in \cite{yrjola2017}. The CBSD (CBRS device, i.e. base station)
 access network and SAS use cases proposed are conceptually similar
to our work. Small start-up cost and growth-aligned transaction costs, 
scalability, exchanges of assets among non-trusting {\it co-opetitive} 
stakeholders, and regulatory reporting and tracking inherent in the 
blockchain model, are argued to be beneficial to CBRS.
Akin to the Cognitive Radio work discussed above, CBRS is mainly
focussed on base stations negotiating spectrum contracts (with SAS)
as opposed to the end-user device, i.e. phone, tablet, laptop etc,
as in our work. Stretching out the market to the deep edge puts
an extra burden on simplifying the interactions, but has the benefit
of increased privacy, as all the context to make contract decisions
can be maintained locally on the device. Like in our case, the CBRS
blockchain analysis recommends hybrid or private blockchains for all
their use cases to reduce transaction time, mainly thanks to simplified
consensus algorithms.

\subsection{Primary Network Spectrum Sharing}
Rahman \cite{rahman2018}, discusses the use case of primary spectrum sharing
among competing operators and proposes a game theoretic algorithm to
avoid free-riding. Simulations show how larger providers can determine which 
proportion of resources to share to avoid siphoning users to smaller providers.
The free-riding problem does not arise in our model since end-users
purchase bandwidth directly in micro-contracts, as opposed to traditional 
roaming agreements. However, similar methods
may be employed in our system by providers to determine the optimal (in terms
of short term and long term revenue) capacity share to give to visiting versus
home-PLMN subscribers.

\subsection{IoT}
Zhou et. al. in \cite{zhou2020} propose to make use of underutilized
human-to-human spectrum frequencies for machine-to-machine use
cases in heterogeneous networks. Privacy, incentive-compatibility
and spectrum efficiency are goals that led the authors to
a blockchain-based solution. In their model, the human users 
relinquish spectrum to the base station in exchange for
payments, and the base station then utilizes this spectrum
for machine-to-machine communication. Blockchain payments
are used to give (human-to-human) users an incentive to
share private information (such as demand) to the base stations to
make better resource allocation decisions. The decentralized
nature of blockchains is also used as an argument for
increased security by not exposing single points of failure.
The two-party market of primary and secondary users is akin
to the cognitive radio approaches. In our work there is
no difference between spectrum users. The base stations
as an extension of their core networks and the operator
simply offers bandwidth that can be purchased by anyone. Then
anyone who can prove they purchased the access grant is let on
as a primary home subscriber on the mobile network. 
Note the operator may
still want to employ something like network slicing or other
constraints to separate long-term subscribers from these
new type of subscribers. Furthermore, in our work the primary users
or subscribers of the network are not involved in the market directly,
but indirectly, where the operator is likely to slice the share 
and set the prices on the market based on their usage and demand at
any given time. The authors also opt for
a permissioned ledger, which they refer to as a consortium
blockchain where the base stations act as authorized nodes,
and miners in the proof-of-work based protocol, to write into the ledger.
A stark difference between their approach and ours is that we consider
a competitive multi-operator environment as opposed to
a single-provider market.

\subsection{Auctions}
Katobi and Bilen in \cite{kotobi2017}, propose an alternative
to the Aloha medium access method to share spectrum more efficiently
amongst competing or interfering cognitive radios.
They design an auction mechanism~\cite{kotobi2016}, as well as introduce a
virtual currency, {\it specoins} that can be earned through mining
to incentivize writing into the blockchain, and thus target
permissionless deployments. In their model, anyone can purchase access on the blockchain and there is no
central entity authorizing either ledger nodes or transaction
participants. The currency is exchanged between primary users
owning the spectrum and secondary users who want to lease it.
Note that all users are cognitive radio base stations in
this context. Apart from earning specoins to afford purchases
of spectrum not owned by mining, base stations may also use
an exchange to trade real currency for specoins.
In order to avoid multiple-round auctions to slow down transactions
and increase the complexity of the exchange, the authors
propose a puzzle mechanism, akin to what is used in the
blockchain ledger update mechanism to assign the winning bidder.
The winning bidder who solves the puzzle is then in a second step 
entitled to pay the required specoins to get the spectrum grant.
Although simulations show that throughput can improve, the fact that
both transaction ledger updates and auction participation
requires processing means that the base station energy and power
consumption may be increased. The spectrum {\it winner} is hence
like in the Aloha model random to some degree.
In contrast, our auction is more like a traditional auction where the highest
bidder wins, or the bid is proportional to the share gained,
in order to help owners of spectrum or bandwidth capacity 
(in our case mobile network operators) determine the
optimal price based on demand. Instead of a single winner
we also allow multiple winners that share the capacity in
resource block shares. Given the random nature of the
auction proposed in \cite{kotobi2017} it may be considered
more of a collision avoidance mechanism than a true open market
auction. The one benefit of their approach compared to
the Aloha protocol is that placing a bid and getting
access requires spending of local resources, in this case
CPUs and in that sense it is harder for a rogue player to
gain access. In our case we rely on the fact that
spending currency is enough of a detractor to gain
rogue access.

\subsection{Access to charging stations for electric vehicles (EVs)}
The fundamentals of a market for a scarce resource, with transactions handled via smart contracts implemented on a blockchain platform, hold for more scenarios than just spectrum.  The use of a blockchain to ensure accurate measurements of the amount of power drawn by a charging electric vehicle (EV) and the accurate billing thereof by the operator of the charging station was considered in~\cite{jeong2018}.  A full marketplace design on a blockchain platform with the operators of fleets of charging stations handling transactions with EV owners who use these charging stations via an application on their smartphones was proposed in~\cite{morat2020}.  Their marketplace for EV access to charging stations is analogous to our proposed marketplace for end-user access to bandwidth, with the owners of the charging stations being analogous to MNOs in our marketplace.  Like in our work, the blockchain implementation in~\cite{morat2020} is also an open-source ledger implementation within the Hyperledger~\footnote{\url{https://www.hyperledger.org/}} program.

\section{Roaming}\label{sec:roaming}
In the US, the FCC mandates that commercial mobile
radio services provide automatic roaming
to other providers on ``commercially reasonable terms and conditions''\footnote{https://www.fcc.gov/wireless/bureau-divisions/competition-infrastructure-policy-division/roaming-mobile-wireless}. In practice this typically means that a provider allows roaming
to other non-competing networks in areas where they do not have a license to
operate, like across national borders.

Below we describe the basic architecture of roaming in a 3GPP
standardized LTE network~\footnote{5G roaming is conceptually similar.}.

An operator of a Public Land Mobile Network (PLMN), also known
as a cellular carrier, has some subscribers that pay the operator for
their service according to (typically long-term) contracts.
The PLMN that the subscriber pays its bills to is typically
referred to as the Home-PLMN. When a subscriber roams
onto another network the new network is referred to as the Visited-PLMN.

The core network (in LTE called EPC, Evolved Packet Core),
is comprised of a set of standardized components to offer
mobile services to subscribers, such as mobility management,
subscriber management, charging, and routing packets to and 
from the Internet. A roaming user would connect to the 
eNodeB (the radio or tower front-end) that is connected to
the EPC of the Visited-PLMN. The mobility management
component in the Visited-PLMN can then interact with the
subscriber data base of the Home-PLMN to verify the
subscription of the user. The Visited-PLMN may connect the
user to the packet data network gateway (P-GW) of the
Home-PLMN to allow the subscriber to access mobile 
services from its home network, while being connected
to the eNodeB of the Visited-PLMN. The P-GW is also typically
responsible for routing packets to and from the Internet.

Now, to charge for
traffic used by a visiting subscriber, the visited
network creates call detail records (CDRs) that are
converted to a Transferred Account Procedures (TAP)
file that is sent to the home network for billing via
a clearing house service.

The home network subscriber verification,
traffic routing, and billing processes will
only proceed successfully if there is a pre-established
bilateral roaming agreement between the 
home network and the visited network. Those
contracts are established manually over long
time spans behind closed doors, and they
are not standardized. Furthermore, they are typically not
established if there is competition in the
same region between the two providers. Hence
traditional roaming cannot be fully utilized
to load balance or fail over between providers 
in the same region, or to allow users a choice
which provider to use at any given time unless
they purchase two home network contracts and
have multiple SIM slots on their phones. In other
words, traditional roaming was designed to
extend coverage rather than maximize performance.

There are new types of mobile virtual network operator
(MVNO) services starting to relax some of these
restrictions, such as GoogleFi~\footnote{\url{https://fi.google.com/about}}, where a custom
core network is built to allow switching between
multiple networks that the MVNO has established
roaming agreements with. Apart from building
a new core network, the manual closed-door
bilateral roaming agreements still need to be set up,
which typically means only a very limited set of
pre-established providers can be used in each region.

\section{Transactions on a spectrum market}\label{sec:blockchains}
\subsection{Smart contracts to handle transactions}
A spectrum market may have an owner of spectrum (i.e., a licensee in the form of an MNO, or maybe the government in the future)  selling access to fine-grained blocks of bandwidth for short durations directly to end-users.  In this scenario, the traditional contracts covering billing and usage may be turned into smart contracts in a straightforward way, so that the smart contracts may now be handled in a fully automated fashion by the blockchain platform.  However, there are other, richer, scenarios that arise when the owners of spectrum do not sell access directly to the end-users and instead sell bandwidth in bulk to an intermediary, which then turns around and sells the bandwidth to end-users.  

In the latter scenario, the intermediary, which we shall call the \emph{aggregator} since it purchases bandwidth in bulk from the spectrum owners, is the service provider to the end-users.  We have already mentioned earlier that end-users are unlikely to want to deal with multiple MNOs directly, thereby providing a niche for such an intermediary, which charges a fixed monthly subscription fee from the end-users who use its services.  Moreover, MNOs do not want to handle the volume of short-duration small-denomination transactions that are required to support lots of end-users directly on such a spectrum market, choosing instead to deal with a few large aggregators who purchase in bulk from these MNOs.  Indeed, we expect the MNOs will even incentivize such aggregators by offering them a volume discount for bulk purchases, coupled with a reservation discount for purchases made in advance of use (the earlier the purchase, the greater the discount).  Although we will not delve into the details here, we will state that smart contracts can handle the transactions between MNOs and aggregators and between aggregators and end-users on such a market.

We note here that the kinds of transactions we have defined above are not the only transactions possible on such spectrum markets.  For example, the MNOs may sell bandwidth access directly to end-users but use aggregator-like entities as sales channels, analogous to hotel chains in the hospitality industry.  In this scenario, the aggregators do not make purchases (either in advance, or in bulk) from the MNOs, but merely send their subscribers (to whom they are the providers of service) directly to the spectrum market to purchase bandwidth access from the MNOs.  As already stated above, this is not attractive to the end-users, as they are unlikely to want to deal individually with multiple MNOs, and nor is it attractive to the MNOs, who need to maintain the ability to sell (and more importantly, provide after-sales service and support) to end-users directly on the market.  (This problem does not arise in the hotel industry because each hotel chain already allows customers to book rooms directly using the reservation system of that hotel chain.) However, smart contracts can handle this scenario as well.

Yet another kind of financial arrangement on such a spectrum market is one where the billing or pricing is not by the bandwidth (MHz), but rather by the throughput when that bandwidth is used (MB).  Although billing based on data usage is the norm in the cellular industry today, and is even seen in the pricing tiers of an advanced MVNO-like service like GoogleFi (\$10 for every 10\,GB after the first 10\,GB), we argue that such billing practices are the consequence of the present illiquid semi-static spectrum allocation mechanism and will not be sustained in a dynamic market for spectrum.  The reason is that the actual scarce quantity is bandwidth and not throughput.  Bandwidth and throughput are related in the sense that large bandwidth allows potentially large throughput, and conversely, large throughput cannot be achieved over a narrow bandwidth.  However, once a certain amount of bandwidth has been assigned to a given end-user for its use, it costs the network operator only negligibly more\footnote{The greater cost comes from increase power consumption to transmit more data.} to support the maximum possible data rate over the allocated bandwidth than to support a lower data rate.  Thus a dynamic market for spectrum will aim to achieve efficient allocation of the actual scarce resource, namely bandwidth, which means that the most direct pricing structure will be that which applies directly to bandwidth rather than indirectly to bandwidth via the throughput when the bandwidth is in use.  Nevertheless, we mention in passing that even this scenario can be handled by smart contracts. 

\subsection{Blockchains to implement smart contracts}
In late 2008 the white paper~\cite{nakamoto2008}, under the presumed pseudonym Satoshi Nakamoto, was distributed on a crypto mailing
list. It proposed a novel peer-to-peer electronic
cash system. Just a few months later Bitcoin was
born based on the algorithms presented.
The algorithm, which today is commonly referred to
as the original blockchain algorithm, describes
a new currency where transactions
are recorded in a decentralized public ledger without
central control from any banking organization.
Anyone may access the ledger and confirm the validity
of both historical transactions as well as the current
latest transaction state. Although the ledger is
public, the identity of the transacting parties
is anonymous. In order to write into the ledger, i.e.
write the next block or transaction into the transaction
history, a distributed consensus protocol is used to ensure
that only a single verified trace of transactions may exist.
In the original Bitcoin paper Proof-of-Work (POW) was used
to ensure this property and to incentivize hosting of the
ledger and moving the ledger forward. POW relies on mining,
or solving complex mathematical problems, where the problems are defined using random hash functions in such a way that their solution necessarily involves trying possibilities one by one until a solution is found. Finding a solution is therefore like winning a lottery, and earns the winner the right to write the next block into the blockchain and in the process
receive a small amount of the cryptocurrency (hence the name ``mining''). 

Since the original algorithm, two flavors of blockchains have
emerged, permissioned and permissionless. In a permissionless
blockchain (like the one in Bitcoin) anyone following the POW rules may
write into the ledger, which is fully public.

In a permissioned blockchain only a pre-defined set of nodes have
access rights to, and may write into, the ledger. Since it is
popular in private Enterprise settings, it is sometimes also called
a private or an Enterprise type blockchain. The main
benefit over the permissionless blockchain is that a permissioned
blockchain may reach consensus quicker and thus increase the rate
of transactions that may be supported.  Moreover, permissioned blockchains may assume that all (or nearly all) nodes are up and running all the time (unlike in permissionless blockchains, where miners may leave or join the network at any time).  This makes it possible for permissioned blockchains to use older Byzantine Fault Tolerance (BFT) consensus protocols which yield a property called \emph{finality} (i.e., non-reversibility of transactions at a certain depth in the blockchain) that is not available in, say, a permissionless blockchain like Bitcoin\footnote{The probability of reversal of a transaction in a block buried deep enough in the Bitcoin blockchain decreases exponentially with the depth of the block, but is never exactly zero.}.

Smart Contracts are a form of self-executing agreement between
buyers and sellers originally proposed in \cite{szabo1994}.
The terms of the agreement are encoded in a computer program
responsible for executing, controlling, and documenting all
events that pertain to the contract. Although this idea predated
the Bitcoin paper by more than a decade, today {\it smart contracts}
, typically and here, refers to programs that are hosted and document their transactions
in a blockchain network.

Non Fungible Tokens (NFTs) are yet another blockchain-related
invention where digital asset ownership as well as transfers
of ownership are recorded on a public blockchain.
The concept was popularized and gained mainstream
prominence in the CryptoKitties game~\footnote{https://www.cryptokitties.co/}.
The verifiably unique asset when minted can define the scarcity and map its
digital version to a physical counterpart. Art, sport video clips, digital 
sport cards and game artifacts have all thrived as NFT assets.

Short transaction execution times, independently verified transactions,
self-contained cryptographically signed standard transaction payloads,
unique assets with flexible scarcity
arrangements, and automatically executing agreements between sellers
and buyers are all properties we borrow from in our Bandcoin proposal.

\section{Design of a blockchain platform for smart bandwidth contracts}\label{sec:design}
This section describes the high level-design of a smart contract
used to sell and purchase mobile network bandwidth contracts 
that grant the holder access to a network on the specified bands with the
specified bandwidth for some given time in a given location.

\subsection{Fundamental Building Blocks}\label{sec:designlocks}
We expect a permissioned blockchain to be used so only certain
trusted or certified actors like MNOs can post offers on
the blockchain. Anyone with a large enough budget may buy bandwidth
according to the rules of the smart contract.

The smart contract is defined in the following terms, all described
in more detail below:
\begin{itemize}
\item{{\bf Actions} what actions are allowed to modify the blockchain ledger.}
\item{{\bf Payload} what fields and values may be submitted to perform certain actions}
\item{{\bf Processor} the processor is where all the {\it smartness} or logic of
the contract is defined. E.g. which actions and what fields and values may be submitted to perform certain actions, and what is the resulting state given a previous state in the ledger.}
\item{{\bf Ledger State} The record that represents the current state of the ledger for a bandwidth offer. The state determines how actions result in new states.}
\end{itemize}

\subsubsection{Transaction Actions}\label{sec:designacitons}
We define the following actions that can be submitted against the bandwidth blockchain:
\begin{itemize}
\item{{\bf create} Both MNOs offering bandwidth and anyone purchasing allocations 
like consumers or service providers must create an account with the public key of their private key being the
unique account name, before performing any transactions. At creation time the account has a 0 balance.}
\item{{\bf deposit} To fund accounts with the bandwidth currency, here called {\it Bandcoins}, an {\it Exchange} needs
to deposit some units of Bandcoins into an account. An Exchange is a trusted {\it super user} on the
blockchain that can ingest currency like a central bank. In practice the Exchange is simply the back-end of a
payment gateway. We will discuss this process in more detail in the implementation section below. }
\item{{\bf withdraw} An MNO will be paid in Bandcoins and may want to exchange that into another currency. The trusted
exchange that is allowed to deposit is also similarly allowed to withdraw funds from the blockchain back to another currency via a payment gateway credit.}
\item{{\bf transfer} Anyone with an account with a non-zero balance, may transfer parts or all of that balance to another
account. Note, for actual purchases and sale of bandwidth funds are transferred atomically so does not have to be done with the transfer action. The transfer action is useful to decentralize funding into accounts. The top-level root exchange may inject
funds into the system, and other accounts may then trickle those funds into individual accounts. There may also
be some out-of band, custom service that one account wants to pay another account for, e.g. a periodic release
of funds like a salary}
\item{{\bf offer} The offer action is used to create a bandwidth offer by a trusted MNO. The offer specifies the bandwidth,
the frequency band, the number of units available for sale, the price per unit, duration of access, and various discounts for bulk purchases and future reservations. Note the offer only has the minimal required meta data about the offer that are contractually binding and written to the ledger. Other meta data may be stored in discovery services. }
\item{{\bf allocate} The allocate action is used to purchase individual units of an offer. If the transaction goes through, e.g. requester has sufficient funds and there are units available for sale, a transaction record will be written to the ledger
that can be used by the receiver to prove its bandwidth grant. The allocate action may also be issued by a 3rd party
such as a service provider to purchase grants for others. In that case, the action would include the public keys of the holders of the private keys that are intended to use the allocations. Note, the private keys may not be mapped to actual consumers at the time of purchase. If no public keys are included, the issuer (transaction submitter) of the action is assumed to be the consumer as well.}
\end{itemize}
 
\subsubsection{Transaction Payload}\label{sec:designpayload}
The fields that need to be transmitted in the payload of a transaction request to the transaction processor depends on
the action performed. But some fields are shared across actions.

\begin{itemize}
\item{{\bf action} one of the actions defined above.}
\item{{\bf provider} In the case of an allocation, deposit, withdrawal the provider is the public key or account name of the target, e.g. the MNO for allocations.}
\item{{\bf price} This is used by all actions except the create action to specify the number of Bandcoins that
pertains to the action. For an offer it would be the unit price. An allocation must specify a price that matches the offer price targeted, as a safety measure for a consumer so they don't get any unsuspecting charges. For deposit, withdrawal and transfer it is the amount involved in the transaction.}
\item{{\bf epoch} Offers are immutable in that each new offer results in a new epoch. The first offer will be given epoch 1. All subsequent offers will bump up the epoch with one. In an allocate request, the consumer or a service provider can hence match the offer listing epoch with the requested contract to make sure they get the posted contract, or their request is reject if a new offer has been posted to replace the old one in the meantime. The epoch field may also be used for future reservations. Even in this case there must be a matching future reservation offer for that epoch for the allocation to succeed.}
\item{{\bf from\_frequency} Lower frequency of the band offered.}
\item{{\bf to\_frequency} Upper frequency of the band offered.}
\item{{\bf bandwidth} Bandwidth offered within the band. Must be less than or equal to {\it to\_frequency} minus {\it from\_frequency}.}
\item{{\bf max\_allocations} Used in the offer action to set the number of units that may be sold (in the current epoch)}
\item{{\bf volume\_discount} A value between 0 and 1 denoting the level of discount for bulk purchases. The $n$th item beyond the first is charged $price \times {volume\_discount}^n$}.
\item{{\bf reservation\_discount} A value between 0 and 1 denoting the level of discount for purchases in advance. Making a purchase $n$
periods in advance costs $price \times {reservation\_discount}^n$. Note that the price can differe between different epochs in the future.}.
\item{{\bf reservations} A dictionary keyed by epoch with the fields, {\it price}, and {\it capacity}. If the epoch exists as a reservation state in the ledger it will be updated, if the capacity is 0, the previous reservation in that epoch will effectively be deleted. All other
cases lead to merges of reservations offered. The capacity is honored for advanced purchases any number of epochs in advance. If the epoch
becomes the current epoch the spot capacity, i.e. what's defined in {\it max\_allocations} becomes the units available for sale.
Note, like with any other types of modification to the offer the epoch is bumped each time a new reservation epoch is added or modified.}
\item{{\bf allocation\_duration} The time a grant is valid, in seconds}
\item{{\bf epoch\_to} In an allocate request the issuer may request a whole series of consecutive future allocations in one transaction. Note,
all epochs must be available for reservation.}
\item{{\bf consumers}. For bulk allocations or for allocation by a 3rd party, public keys may be specified for who is allowed to consume
the allocation. The private keys may be mapped to actual users after the transaction has completed.}
\end{itemize}

\subsubsection{Transaction Processor}\label{sec:designprocessor}
The transaction processor defines how a current ledger state is transformed into a new state and what payloads are needed
for the transaction to succeed. The transaction processor also cryptographically verifies the issuer of the transaction
using PKI. For instance, an offer can only be posted by the account owner as per the public key of the issuer.
We only list the rules for the offer and allocate actions here as the other actions just perform trivial transfers between
accounts as you would expect from any banking like transaction. An account balance cannot go below 0 as a general rule.
In terms of the transfer the verified issuer needs to match the account the funds are transferred from.
\begin{itemize}
\item{{\bf offer} all fields mentioned above need to be specified except provider (assumed to be issuer), epoch(\_to) as it is bumped
up by the processor, and consumers (as it is for the purchaser to restrict who can access the grant, with the exception of auctions as we will see below). In practice only trusted MNOs would be
allowed to post offers. Note that setting an offer with max\_allocations to 0 is essentially turning off the offer and not allowing
any more allocations. Whenever an offer is posted, the allocations left in the state (see below) is reset to max\_allocations}
\item{{\bf allocate} An allocation must specify a provider target that must match an offer. The price and band info must also match the
current offer as a safety check, but most crucially the epoch must match. The consumers field may be filled out to delegate the
allocation to someone except the issuer, and should also be used for bulk purchases. In the case of reservation and advanced purchases, the
price needs to match the total computed cost for the full purchase not the spot price in the offer. This again verifies that the issuer
and the transaction processor are using compatible pricing algorithms and there are no surprises in what is charged to the issuer account.
The issuer account will get this amount withdrawn if there are sufficient funds and the target account will be credited with this amount.
The number of allocations requested needs to be equal to or less than allocations left in the ledger state.}
\end{itemize}

\subsubsection{Ledger State}\label{sec:designledger}
On completion of a successful transaction the ledger state is updated and the payload that was submitted is recorded in the transaction log
atomically and with consensus verification in the blockchain.
The following state is recorded and available for verification of current offers and account balances:
\begin{itemize}
\item{{\bf name} public key in hex of the account holder.}
\item{{\bf balance} current account balance in Bandcoins.}
\item{{\bf allocations\_left} allocation units that are left in current offer.}
\item{{\bf epoch} current epoch for current offer.}
\item{{\bf from\_frequency} Lower frequency band in current offer.}
\item{{\bf to\_frequency} Upper frequency band in current offer.}
\item{{\bf bandwidth} bandwidth of current offer.}
\item{{\bf price} (spot) price of current offer.}
\item{{\bf volume\_discount} volume discount of current offer.}
\item{{\bf reservation\_discount} reservation discount of current offer.}
\item{{\bf allocation\_duration} allocation duration of current offer.}
\item{{\bf reservations} dictionary keyed by epoch of future reservations that may currently be purchased.}
\end{itemize}
Note, only {\it name} and {\it balance} are relevant for a consumer account that is not also
a provider.

\subsection{Interacting with the Blockchain}\label{sec:designinteract}
Given that the payloads, processing, and state are all defined, the processors may run and be hosted
by any and all nodes making up the blockchain infrastructure. Below we will show the basic means by
which external parties interact with this blockchain.

\subsubsection{Transaction Client}\label{sec:designclient}
A transaction client library is provided capable of packaging transaction
requests with the correct payloads for the actions available. It also handles
local key management and signing, encapsulation and serialization of transactions.
Transactions may be submitted in batches by 3rd parties, i.e. does not have to
be submitted by the entity signing individual transactions in the batch.
Offer creation may be restricted to authorized users such as trusted
and certified MNOs. The transaction client also uses blockchain APIs
to retrieve the current state of accounts, including account balances
and current offers, as well as listing transactions for verification.

\subsubsection{Transaction Tunneling}\label{sec:designtunneling}
Given that all transaction requests are signed by the issuer,
the requests may be tunneled easily in routers
and transported through different 3rd parties such as service providers
or auctioneers as we shall see below. We may also tunnel the transactions
through RF links such as LTE NAS. The transactions when executed
are logged in the public ledger together with the resulting state
for anyone to verify, so there is no inherent need to encrypt the payload.
Transaction replay is not possible by virtue of the payload signature design,
and should multiple transaction processor try to execute the request concurrently
only one will succeed to write into the consensus approved ledger.
However, a 3rd party or middleman may delay or stop a transaction request from
being executed, again something that is leveraged in the auction design
below.

\subsubsection{Payment Gateway Exchange}\label{sec:designexchange}
As we alluded to in the actions section above deposits and withdrawals
are used to exchange other currencies for Bandcoins in the blockchain.
Trusted exchanges will integrate with a payment system in the form of an
online store or a payment server, which upon verification of payment
from another currency will deposit funds into the bandwidth blockchain.
The reverse is also possible where the exchange can pay into another
currency if the requester can prove they own a Bandcoin account in the blockchain,
e.g. if a MNO wants a payout in traditional currency.
Each exchange will thus also define the exchange rate. One could also
imagine deposits being made as an incentive of an external action, such
as signing up for a subscription for another service, and of course 
an Exchange could give bulk discounts for transferring Bitcoins
into the blockchain as well, like minutes on a call plan.
Yet another use case is some service provider that decides to transfer
credits into its subscribers' accounts on a periodic basis or as needed.

\subsection{Blockchain consensus protocol design}
Unlike the blockchain networks for public cryptocurrencies like Bitcoin and Ethereum (which was designed to support smart contracts), the ``nodes'' running the blockchain network that supports the smart contracts for the proposed spectrum market may be assumed to have cleared scrutiny and hence be \emph{trusted}.  This is because the nodes are components of a cellular wireless network where such authentication, validation, and enforcement/checking of trust are built into the existing system.  

Recall, a blockchain network running on a vetted set of nodes is called a permissioned blockchain (as opposed to Bitcoin, for example, which is an example of a permissionless blockchain network).  The use of a permissioned blockchain does not eliminate the need for security -- in particular, the property of Byzantine Fault Tolerance (BFT) to ensure security in the cryptographic sense against an intelligent adversary who may hack/hijack/bribe a certain fraction of the nodes.  However, the permissioned blockchain allows for the use of consensus protocols like PBFT~\cite{castro1999} that also provide the benefits of \emph{finality} (i.e., a smart contract cannot be voided in the future by an adversary forcing the blockchain to switch to a different fork that does not contain the block with that smart contract) along with low latency (i.e., a smart contract can be finalized quickly), neither of which holds for a permissionless blockchain like Bitcoin.  Low latency is important because smart contracts involving allocation of bandwidth blocks to end-users need to be finalized before the epoch when that bandwidth block will be used by the end-user.  

\subsection{Auction Design}\label{sec:designauction}
With the basic smart contract in place, providers and consumers can exchange
access right, but one obstacle remains for providers to participate:
-how should the price be set to meet the demand?

Here, we propose a mechanism that integrates the trust features of a distributed 
blockchain with the functionality of a clearing house you would normally see in a traditional auction.

The blockchain provides distributed smart contracts that may be verified by any party at the time of provisioning the resources, and it also takes care of atomic monetary transfer of funds between consumers and providers. The auction clearing house is responsible for soliciting prospective consumer demand by mediating pending blockchain transactions and finding matching bids and offers, and ultimately setting the clearing price and executing the winning transactions.

In terms of the basic blockchain mechanism we add one component to the ledger state, a {\it consumers}
dictionary with public keys of consumers with the following fields:
\begin{itemize}
\item{{\bf allocated} a boolean indicating whether the consumer has claimed this allocation}
\item{{\bf allocations} number of allocations granted to the consumer}
\item{{\bf price} the price the consumer must pay for the allocation}
\end{itemize}

The network provider will pass in this dictionary in an offer to indicate a settled auction
allocation and may assign different shares and prices to different consumers.
Only consumers with matching keys in this dictionary may then purchase the allocations specified
with the given price.

\subsubsection{Auctioneer Clearing House}\label{sec:designauctioneer}
The clearing house is intended to mediate between entities that offer and purchase smart contracts on the blockchain. If the blockchain is used directly by consumers and providers it could be seen as a posted-price spot or on-demand market. Such markets may be efficient when providers have enough information to set the price that correctly match supply with demand, but soliciting such information with varying offers over time would be inefficient, in particular in the use case of interest of automated short-term micro-contracts for cellular spectrum resources.

A feature of blockchain transactions is that the input may be cryptographically signed and put on hold before executed by a 3rd party against the transaction processor for ledger recording and verification. That means that we can submit the purchase transactions before the offer transactions and keep them in a pending state in the clearing house until a price is determined, at which time an offer is generated and the winning bid(s) is(are) executed as a normal purchase transaction(s).

By allowing providers insight into the pending bids, they may set an offer price that is more in line with the current demand. Conversely, the consumer could reveal their true preferences in a sealed bid without being taken advantage of with higher-than-demand pricing.

For different clearing mechanisms, what information is revealed and exposed to the consumer (bidder) and provider (offer generator) may vary. We give examples of Vickery and Proportional share auctions below.

The consumer must however be able to, at a minimum, query the state of their bid to be able to determine if they had a winning bid that got executed, so they can exercise the allocation and use the cellular network bandwidth purchased.

The successful execution of a transaction submitted by a consumer (as a bid) relies on the provider creating an offer that matches and that is only possible to purchase by that one consumer.

In case of second bid auctions this may provide a challenge as the final price is not the price the consumer agreed to pay but the price submitted by the second highest bid. In that case we propose a price refund mechanism, where the first price transaction is submitted and the provider then refunds the consumer the difference.

The clearing house needs to support the protocol in Table~\ref{T:clearinghouseapi}
and the provider the protocol in Table~\ref{T:providerapi}.
The provider protocol is needed so the clearing house can ask the provider
to sign a cleared offer configuration into an offer on the blockchain.
The clearing house will then execute the offer transaction on behalf of the provider,
so the provider does not need to have direct access to the blockchain. This design
would allow delegation of granting different trusted providers access to enter offers.
The auctioneers need to be trusted and then whoever they trust as providers may effectively
also be trusted on the blockchain.

\begin{table*}[htbp]
        \caption{Clearing house protocol.}
\begin{center}
\begin{tabular}{|l|l|l|l|}
\hline
\textbf{CreateAuction} & & & The operation is invoked by the provider to create auctions. \\
 & & & \\
 & IN & auction\_type & $[VICKREY|PROPSHARE]$ \\
 & IN & max\_allocations & max allocations of offer \\
 & IN & provider & public key of signer of offer \\
 & IN & clearing\_time & absolute time in epochs when the auction should be cleared \\
 & & & and winning bid(s) executed \\
 & IN & from\_frequency & offer from\_frequency \\
 & IN & to\_frequency & offer to\_frequency \\
 & IN & bandwidth & offer bandwidth \\
 & IN & epoch & current offer epoch + 1 \\
 & IN & min\_price & all bids need to be at least this price \\
 & IN & offer\_callback & will be called by clearing house when auction is cleared to \\
 & & & generate offer and refund transactions \\
 & OUT & auction\_id & uniquely generated auction ID \\
\hline
\textbf{ListAuctions} & & & The operation may be invoked both by the consumer to show \\
 & & & details about auctions to place a bid, or for providers to get \\
 & & & information about bids. Bids may or may not be revealed \\
 & & & depending on whether it is a sealed bid auction or not. \\
 & & & \\
 & IN & auction\_id & optional to filter the results by auction\_id \\ 
 & for each listing: &  & \\ 
 & OUT & auction\_id & auction id \\ 
 & OUT & offer\_details & info from CreateAuction call \\ 
 & OUT & bid\_details & info about current bids (unless sealed) \\ 
 & OUT & status & $[ACTIVE|INACTIVE]$ \\ 
\hline
\textbf{Bid} & & & The operation is invoked by the consumer. It embeds a \\
 & & & blockchain allocation request.\\
 & & & \\
 & IN & auction\_id & auction ID to bid is for\\ 
 & IN & bid & encoded serialized and signed blockchain allocation including \\
 & & & bid price \\
 & OUT & status & $[PENDING|EXECUTED|CANCELLED|ERROR]$ \\
 & OUT & bid\_id &  public key used to sign bid\\
\hline
\textbf{GetBid} & & & The operation is invoked by the consumer to retrieve status \\
& & & of bid.\\
 & & & \\
& IN & auction\_id & auction ID bid is for\\ 
& IN & bid\_id & bid ID\\ 
 & OUT & status & $[UNKNOWN|PENDING|EXECUTED|$ \\
 & & & $CANCELLED|ERROR]$ \\
\hline
\end{tabular}
\label{T:clearinghouseapi}
\end{center}
\end{table*}

\begin{table*}[htbp]
        \caption{Provider (callback) protocol.}
\begin{center}
\begin{tabular}{|l|l|l|l|}
\hline
\textbf{GenerateOffer} & & & Used by the clearing house to generate offers and refunds \\
& & & while clearing an auction. \\
 & & & \\
& IN & price & offer price \\ 
& IN & refund & refund amount \\ 
& IN & from\_frequency & offer from\_frequency \\ 
& IN & to\_frequency & offer to\_frequency \\ 
& IN & bandwidth & offer bandwidth \\ 
& IN & consumers & Consumer specific allocations and pricing to be put in offer (optional) \\
& & & to restrict who can make purchases \\ 
& IN & max\_allocations & offer max allocations \\ 
& OUT & offer & encoded serialized and signed blockchain offer transaction \\ 
& OUT & refund & encoded serialized and signed blockchain transfer transaction \\ 
\hline
\end{tabular}
\label{T:providerapi}
\end{center}
\end{table*}

Next, we detail two examples of using this protocol for running two different kinds of auctions:
{\it Vickrey} auctions and {\it Proportional Share} auctions.
\subsubsection{Vickrey Auctions}\label{sec:designvickrey}
\begin{figure}[htbp]
        \centerline{\includegraphics[scale=0.23]{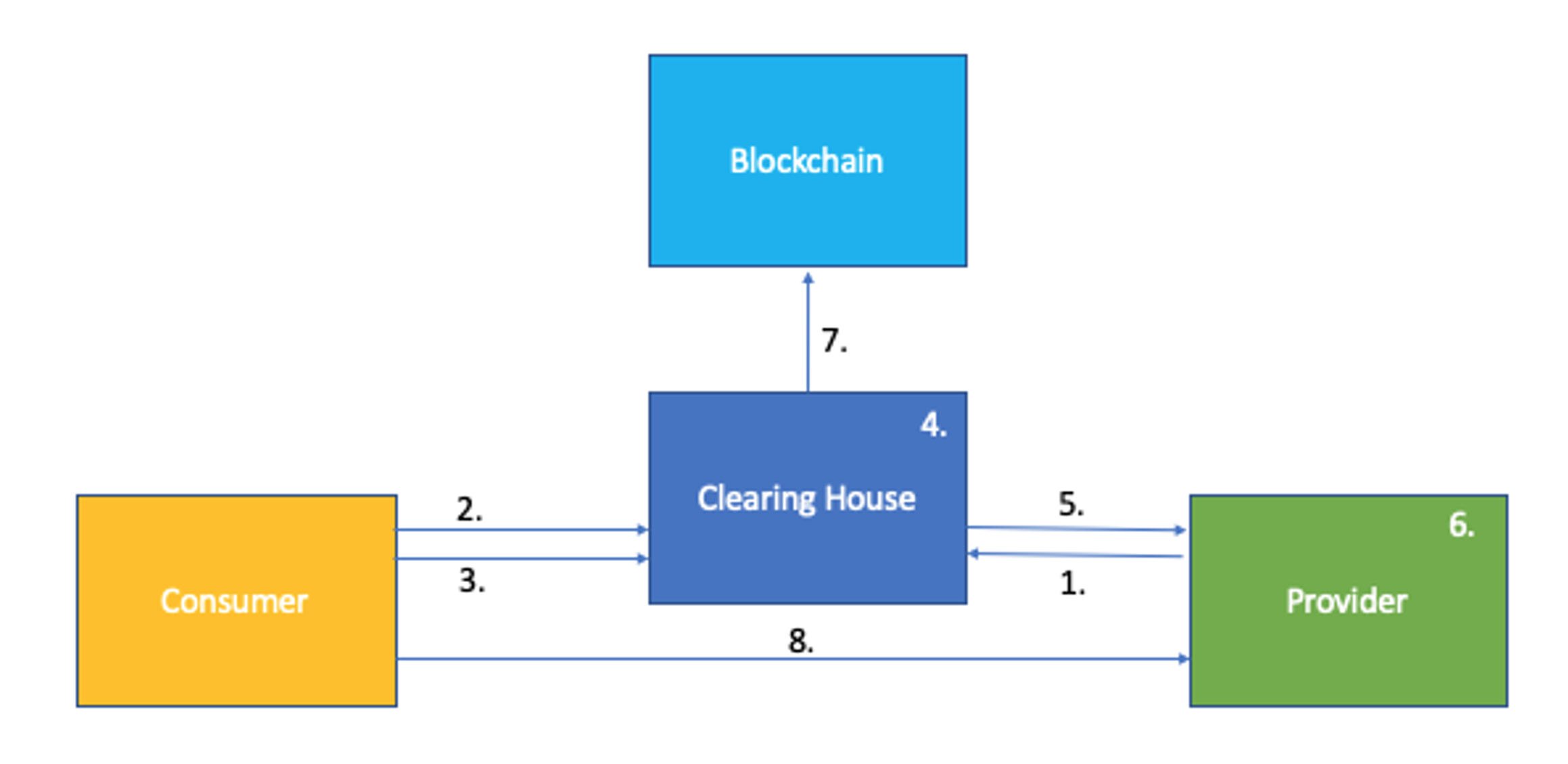}}
        \caption{Vickrey Auction interactions.}
\label{vickrey}
\end{figure}

The Vickrey Auction~\cite{vickrey1961} is an auction where the bidders submit their bids without knowing the bids of others in the auction, and the highest bidder wins but the price is the second highest bid.
The interactions are depicted in Figure~\ref{vickrey}

\begin{enumerate}
\item{Provider creates an auction with type VICKREY to ensure bids are sealed.}
\item{Prospective Bidders get information about the bid with ListAuctions and GetAuction operations.}
\item{The Consumer who wants to submit a bid creates an allocate Transaction that can be executed on the blockchain with its bid price targeted for the provider Offer (which does not exist yet)}
\item{The clearing house verifies that the bid is valid (can be executed on the blockchain)}
\item{After the clearing\_time of the auction passed the auction will stop accepting bids and the clearing house will call the offer\_callback requesting an offer to be created with the winning bidder(s) as the targeted purchaser in the consumers dictionary. The offer will have the winning bid price, but a refund transaction is also requested with the amount of the winning bid – second highest bid.}
\item{The provider will generate offer and transfer blockchain requests matching the info from the clearing house.}
\item{The clearing house will execute the offer transaction, the allocate transaction(s) of the winning bidder(s), and the refund transaction in that order and update the winning bid(s)’ status to EXECUTED. All other losing bids will have their status changed from PENDING to CANCELLED.}
\item{The consumer will query its bid and note it got executed at which point it may request the spectrum allocation from the provider and query the transaction for proof from the blockchain.}
\end{enumerate}

\subsubsection{Proportional Share Auctions}\label{sec:designpropshare}
A {\it Proportional Share} auction~\cite{waldspurger1995} is an auction where each bidder receives a share of the resource
proportional to the bid over all other bids on the same resource. It is commonly used
as an alternative to second bid sealed-bid auctions as it is efficient to implement and known to
converge quickly in practice.

Only steps 1 and steps 6 change compared to the Vickrey auction example above.
In step 1 the auction type is set to PROPSHARE.
In step 6 the following processing is done to clear the auction:
\begin{enumerate}
\item{The sum of all bids is computed - > bid\_sum}
\item{For each bidder an allocation size is computed as $bid\_price \times \frac{max\_allocations}{bid\_sum}$}
\item{A consumers dictionary is created with all bids with at least 1 allocation in step 2, setting the allocations to the computed value in step 2 and the price to the price bid by that bidder}
\item{The offer\_callback is called to generate the offer}
\item{The offer transaction received from the callback is executed}
\item{If the offer transaction is successful execute all bids with an allocation greater than 0}
\item{Change the status of all successfully executed transactions to EXECUTED and the others to CANCELLED (if allocations is 0) or ERROR (if the transaction failed)}
\end{enumerate}

\section{Implementation Notes}\label{sec:implementation}
We have implemented the design proposed in the previous
section using the {\it Hyperledger Sawtooth Core}~\footnote{https://www.hyperledger.org/use/sawtooth} blockchain.

The blockchain state as well as the payloads are encoded as JSON objects.
Our implementation interacts with {\it Sawtooth} mainly through the REST API,
via SDKs. One of the key reasons we picked Sawtooth, apart from having an easy-to-deploy
development environment based on docker, was the multi language support. 
It was easy to integrate with our Python backends, as well as our Android clients.

We provide a Transaction Processor that registers with the validator, and
a client library, both implemented in Python3. All our services
are implemented with Python Flask REST APIs using JSON payloads as well.

The tunneled transactions use Protocol Buffer~\footnote{https://developers.google.com/protocol-buffers} serialized {\it Transaction} and {\it Batch} encapsulations according to the Sawtooth protocol~\footnote{https://sawtooth.hyperledger.org/docs/core/releases/latest/ \_autogen/txn\_submit\_tutorial.html} and are then base64
encoded in JSON fields for our APIs. 

All sawtooth and custom services are hosted in Docker containers and interact within a private
docker network. The services may be distributed with something like Docker Swarm or Kubernetes,
but for testing they all run locally using docker-compose and the developer mode blockchain.

The transaction processor implementation allows a public key to be set with Exchange privileges.
That means that whoever can prove they have the private key corresponding to the
public key may perform the privileged operations such as injecting or removing funds
with the {\it deposit} and {\it withdraw} actions.
The service that owns the Exchange private key is the point of integration with payment gateways.
We have implemented two integrations, Saleor and Braintree, discussed next.

\subsection{Saleor Payment Integration}\label{sec:implementationsaelor}
Saleor~\footnote{https://saleor.io/} is an open-source, customizable e-commerce
shopping cart solution, where popular credit cards can be used to pay for items.
We added custom product items that allows you to purchase Bandcoins in different
chunks, such as 100, 200, 500, 1000 Bandcoins at different USD prices.
At checkout of any of these items the public key of the blockchain account
you want to transfer Bandcoins into has to be specified. The person making
the payment does not need to own the private key necessarily as it can fund
anyone's account with this mechanism. As the order goes from pending to cleared
by the payment system, our Exchange service checks the Saleor order database to
detect new finalized orders where payments have been confirmed. When the order is
deemed finalized, the Exchange service pulls the number of Bandcoins specified in
the order and creates and executes a {\it deposit} transaction with that amount
on the blockchain.

Saleor supports a dummy payment provider that is convenient for development
but also many other providers through payment gateways, including Braintree (see below).

Saleor is a great tool for getting started quickly with a Web portal to 
fund blockchain accounts with traditional currency. The process of entering 
the public key to fund and going through all the steps of searching for
product and adding them to the shopping cart is however a bit tedious,
in particular from mobile phones where you may want to perform fund
operations to be able to purchase mobile access. Therefore we also
provide a mobile API implemented in Android, using the Braintree
payment gateway system, described below.

\subsection{Braintree Payment Integration}\label{sec:implementationsbraintree}
Braintree~\footnote{https://www.braintreepayments.com/} is a payment gateway service
that handles integration with most popular payment providers such online wallets, 
credit and debit cards. 

Its Android API is very easy to integrate into your own app with a few calls to a
service you need to implement to handle payment processing. The service needs to be registered
in your Braintree account online, but then a python Flask REST API server can easily be implemented to interact with your app and Braintree. This server runs in our Exchange service and will
on completion of payment processing credit the blockchain account of the Android phone
with the corresponding number of Bandcoins that was bought.

This allows the phone to easily replenish Bandcoins spent without leaving the app. For the end-user it also provides a one or two-click solution, as a credit card that has been used once can be reused from the same app with a single click to buy additional Bandcoins. 

The backend of this solution is also more reliable than the Saleor integration as our
custom payment server can get notified directly when a payment has been completed and does not
have to poll orders for status.

\subsection{QoS Extensions}\label{sec:implementationextensions}
The payload structure for allocations and offers is encoded in JSON and thus lends itself well
to extensions and customizations. One set of extensions that may be of interest is QoS levels.
The provider may post different offers at different prices and with different custom QoS levels,
which later can be enforced in the mobile core, for example by using network slicing or usage accounting.
Many mobile plans today are not only restricted by access time but also capped by data usage. So we could add in
a GB limit as a QoS level, and be able to charge more for the same contract with a higher cap. 
We note, though, that capping in this way is less critical than in traditional contracts given that
they are by design only intended for short periods of time when the customer actually wants to use the
network.

Similarly, some post-payments may be desirable for the providers, based on usage, i.e. if that data cap
is exceeded. This feature could again be specified in a contract extension and enforced post-contract based
on accounting information as well as making use of the {\it transfer} blockchain action to pay for additional
services. Again, shorter time spans, and the ability for consumers to simply switch to another provider if
QoS is not met, makes these post-payment plans less fundamental than in a traditional roaming setup. 

To allow for these extensions to be passed into offers and recorded on the blockchain ledger they could be encapsulated
in a well-known {\it qos} dictionary field.

\subsection{Transaction Processing Performance}\label{sec:implementationtransactions}
\begin{figure}[htbp]
        \centerline{\includegraphics[scale=0.5]{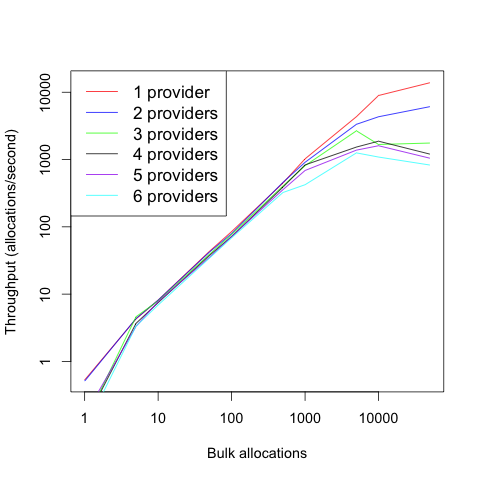}}
        \caption{Transaction throughput with different bulk sizes and concurrent providers.}
\label{transactions}
\end{figure}

Transaction processing time both in terms of latency and throughput matters when it comes to the granularity
of contracts we can support in the time domain. The original blockchain protocol is notorious for
being slow at confirming transactions, and the cryptographic routines used to secure all ledger
writes also bring overhead. Sawtooth offers some optimizations in terms of transaction parallelism
but if you are writing to the same state the parallelism that is possible is limited. As a result, we
introduced bulk allocations, where multiple allocations to different consumers, distinguished by
different public keys, may be purchased in a single Sawtooth transaction. See the design section
for details on how this is implemented in the protocol and the smart contract.
To test the limits and the improvement of bulk allocations we tested different bulk sizes and different
concurrent provider purchases with the same consumer account.
The results can be seen in Figure~\ref{transactions}.

In summary, we can improve the throughput of allocation agreements settled by four orders of magnitude
(from about .5-1 allocations per second to 10,000-13,000 transactions per second) for a single provider with a bulk size
up to 50,000. We see that bulk sizes over 10,000 do not improve performance much but that there is a
linear increase for bulk sizes 1-10,000. Submitting allocation transactions for multiple providers
concurrently is not as efficient of a mechanism to improve throughput as bulk allocations, as seen
by the fact that 3-6 parallel transactions (providers) only reaches a peak throughput of about 1000
allocations per second. The drop of higher parallelism with small bulk sizes could be seen as an experiment setup artifact,
where previous large bulk sizes impact the restart of the experiment run. It also highlights the general strain
on the system of increased transaction parallelism. Another interesting
phenomenon is the drop at the high end, where throughput is satisfied, which is clearly a function of
the number of parallel transactions, and again points to the efficiency of our bulk allocation
mechanism.

\section{Lessons Learned}\label{sec:lessons}
\begin{table}[htp]
        \caption{Comparison of Bandcoin transactions with other technologies.}
\begin{center}
\begin{tabular}{|l|c|c|c|c|}
\hline
 & \textbf{Bandcoin} & \textbf{Roam} & \textbf{NFT} & \textbf{Bitcoin} \\ 
\hline
\textbf{Auto} & X & & X & X \\
\textbf{Fast} & X & & & \\
\textbf{Unique} & X & X & X & \\
\textbf{Periodic} & X & X & & \\
\textbf{Scarcity} & X & X & X & \\
\textbf{Bulk} & X & & & \\
\textbf{Reserve} & X & & & \\
\textbf{Tunnel} & X & & & \\
\textbf{Auction} & X & & & \\
\hline
\end{tabular}
\label{T:summary}
\end{center}
\end{table}

We have found that the permissioned blockchain model
to be a very powerful
comcept to automate bandwidth contract purchases
in mobile networks. A contract can be purchased
within a second~\footnote{local host full RPC from transaction submission to
ledger reporting committed as 0.5-1s in our test, with single validator devmode consensus algorithm.} assuming funding has already been setup and made available through a payment gateway
as described in the previous section. Note that transaction
throughput may be improved by making use of transaction batches
or purchasing allocations for multiple users at a time, or multiple
blocks at a time.
The sub second contract negotiation is easily
hidden in the process of switching network providers
as the LTE Attach call alone to authenticate
takes close to a second in most networks, and as a comparison
loading an eSIM on a phone takes 2-3 seconds.
This mechanism opens the door for much shorter
and finer grained contracts for mobile access including
roaming agreements.
Traditional physical SIM contracts are typically negotiated
on a yearly basis and eSIMs today are commonly offered
on a monthly or weekly basis. The contracts described
here could realistically be offered down to a 10-15 second basis,
but more realistically on a minute-by-minute or hour-by-hour basis.
These times can be cut further when phones start offering multiple
eSIM slots to make it even faster to switch between providers.

The cryptographic guarantees and the message-level as opposed to
transport-level security offered by the blockchain design
is also ideal for service provider integrations, similar to
our auction service design, where a third party service
purchases allocations on behalf of its users.

We have also experimented with tunneling these contract negotiations
with the LTE protocol using the SIB and NAS mechanisms that can be run in a disconnected
state before carrier authentication. Some more details on that work
are available in ~\cite{sandholm2021}.

NFT assets share many features to our bandwidth contracts.
Similarities include, ability to mint scarcity and asset price 
at contract setup, asset sharing, unique and distinguishable (enforced through our epoch mechanism) 
asset tracking and 3rd party verification, and maintaining 
blockchain records of all transactions.We don't support blockchain recorded transfer 
of ownership and royalty
processing as in NFTs, instead we allow external resale of grants
encoded as PKI keys~\footnote{A design critical to our bulk allocation performance improvements.}. Furthermore we support an inherent bidding model
where prices are determined in auctions and the bids may be kept private
and result in proportional grants of resource allocations.

We have summarized the differences in Table~\ref{T:summary}, where {\it roam}, denotes
the current LTE roaming protocol, {\it unique} and {\it periodic} pertain to
the issued grants, {\it scarcity} refers to whether you have control over
the scarcity of resources at creation or minting time, {\it bulk} and {\it reserve} 
indicate whether bulk allocations and advanced reservation of resource blocks are 
supported, respectively. Finally, {\it tunnel} refers to whether 3rd parties can execute
encapsulated transactions, and {\it auction} indicates whether auction purchases
(demand-discovery) is supported.

\bibliographystyle{plain}
\bibliography{related}

\end{document}